\documentclass{wscpaperproc}
\usepackage{latexsym}
\usepackage{graphicx}
\usepackage{mathptmx}

\usepackage{amsmath}
\usepackage{amsfonts}
\usepackage{amssymb}
\usepackage{amsbsy}
\usepackage{amsthm}

\usepackage[pdftex,colorlinks=true,urlcolor=blue,citecolor=black,anchorcolor=black,linkcolor=black]{hyperref}

\newtheoremstyle{wsc}
{3pt}
{3pt}
{}
{}
{\bf}
{}
{.5em}
{}

\theoremstyle{wsc}

\begin{document}

%
%

\pagestyle{fancyplain}

\thispagestyle{plain}
\firstPageHead{}

\chead{\fancyplain{}{\itshape Chopra, Gel, Subramanian, Krishnamurthy, Romero-Brufau, Pasupathy, Kingsley, and Raskar}}

\rhead{}
\cfoot{}
\renewcommand{\headrulewidth}{0pt} 

\makeatletter
\let\@internalcite\cite
\def\cite{\def\@citeseppen{-1000}%
    \def\@cite##1##2{(##1\if@tempswa , ##2\fi)}%
    \def\citeauthoryear##1##2##3{##1 ##3}\@internalcite}
\def\citeNP{\def\@citeseppen{-1000}%
    \def\@cite##1##2{##1\if@tempswa , ##2\fi}%
    \def\citeauthoryear##1##2##3{##1 ##3}\@internalcite}
\def\citeN{\def\@citeseppen{-1000}%
    \def\@cite##1##2{##1\if@tempswa, ##2)\else{}\fi}%
    \def\citeauthoryear##1##2##3{##1 (##3)}\@citedata}
\def\citeA{\def\@citeseppen{-1000}%
    \def\@cite##1##2{(##1\if@tempswa , ##2\fi)}%
    \def\citeauthoryear##1##2##3{##1}\@internalcite}
\def\citeANP{\def\@citeseppen{-1000}%
    \def\@cite##1##2{##1\if@tempswa , ##2\fi}%
    \def\citeauthoryear##1##2##3{##1}\@internalcite}
\def\shortcite{\def\@citeseppen{-1000}%
    \def\@cite##1##2{(##1\if@tempswa , ##2\fi)}%
    \def\citeauthoryear##1##2##3{##2 ##3}\@internalcite}
\def\shortciteNP{\def\@citeseppen{-1000}%
    \def\@cite##1##2{##1\if@tempswa , ##2\fi}%
    \def\citeauthoryear##1##2##3{##2 ##3}\@internalcite}
\def\shortciteN{\def\@citeseppen{-1000}%
    \def\@cite##1##2{##1\if@tempswa, ##2\else{}\fi}%
    \def\citeauthoryear##1##2##3{##2 (##3)}\@citedata}
\def\shortciteA{\def\@citeseppen{-1000}%
    \def\@cite##1##2{(##1\if@tempswa , ##2\fi)}%
    \def\citeauthoryear##1##2##3{##2}\@internalcite}
\def\shortciteANP{\def\@citeseppen{-1000}%
    \def\@cite##1##2{##1\if@tempswa , ##2\fi}%
    \def\citeauthoryear##1##2##3{##2}\@internalcite}
\def\citeyear{\def\@citeseppen{-1000}%
    \def\@cite##1##2{(##1\if@tempswa , ##2\fi)}%
    \def\citeauthoryear##1##2##3{##3}\@citedata}
\def\citeyearNP{\def\@citeseppen{-1000}%
    \def\@cite##1##2{##1\if@tempswa , ##2\fi}%
    \def\citeauthoryear##1##2##3{##3}\@citedata}
%
%
%
\def\@citedata{%
    \@ifnextchar [{\@tempswatrue\@citedatax}%
                  {\@tempswafalse\@citedatax[]}%
}

\def\@citedatax[#1]#2{%
\if@filesw\immediate\write\@auxout{\string\citation{#2}}\fi%
  \def\@citea{}\@cite{\@for\@citeb:=#2\do%
    {\@citea\def\@citea{, }\@ifundefined
       {b@\@citeb}{{\bf ?}%
       \@warning{Citation `\@citeb' on page \thepage \space undefined}}%
{\csname b@\@citeb\endcsname}}}{#1}}%

%
\def\@citex[#1]#2{%
\if@filesw\immediate\write\@auxout{\string\citation{#2}}\fi%
  \def\@citea{}\@cite{\@for\@citeb:=#2\do%
    {\@citea\def\@citea{; }\@ifundefined
       {b@\@citeb}{{\bf ?}%
       \@warning{Citation `\@citeb' on page \thepage \space undefined}}%
{\csname b@\@citeb\endcsname}}}{#1}}%

%
\def\@biblabel#1{}
\makeatother



\newdimen\bibindent
\bibindent=0.0em
\def\thebibliography#1{\section*{\refname}\list
   {}{\settowidth\labelwidth{[#1]}
   \leftmargin\parindent
   \itemindent -\parindent
   \listparindent \itemindent
   \itemsep 0pt
   \parsep 0pt}
   \def\newblock{}
   \sloppy
   \sfcode`\.=1000\relax}


\setlength{\baselineskip}{12.7pt}

\title{DeepABM: SCALABLE, EFFICIENT AND DIFFERENTIABLE AGENT-BASED SIMULATIONS VIA GRAPH NEURAL NETWORKS}

\author{Ayush Chopra\\
Ramesh Raskar\\[12pt]
Media Lab\\
Massachusetts Institute of Technology \\
75 Amherst St \\
Cambridge, MA 02139, USA
\and
Esma S. Gel\\[12pt]
School of Comp., Inf. and Decision Systems Eng. \\
Arizona State University\\
699 S Mill Avenue \\
Tempe, AZ 85281, USA
\and
Jayakumar Subramanian\\
Balaji Krishnamurthy\\ [12pt]
Media and Data Science Research Lab \\
Adobe\\
Block A, Sector 132 \\
Noida, Uttar Pradesh 201304, INDIA
\and
Santiago Romero-Brufau\\ 
Kalyan S. Pasupathy\\[12pt]
Department of Healthcare Systems Engineering\\
Mayo Clinic \\
200 1st St SW \\
Rochester, MN 55905, USA
\\
\and
Thomas C. Kingsley\\ [12pt]
Alix School of Science and Medicine\\
Mayo Clinic \\
200 1st St SW \\
Rochester, MN 55905, USA
}

\maketitle

\section*{ABSTRACT}
We introduce DeepABM, a framework for agent-based modeling that leverages geometric message passing of graph neural networks for simulating action and interactions over large agent populations. Using DeepABM allows scaling simulations to large agent populations in real-time and running them efficiently on GPU architectures.
To demonstrate the effectiveness of DeepABM, we build DeepABM-COVID simulator to provide support for various non-pharmaceutical interventions (quarantine, exposure notification, vaccination, testing) for the COVID-19 pandemic, and can scale to populations of representative size in real-time on a GPU. Specifically, DeepABM-COVID can model 200 million interactions (over 100,000 agents across 180 time-steps) in 90 seconds, and is made available online to help researchers with modeling and analysis of various interventions. We explain various components of the framework and discuss results from one research study to evaluate the impact of delaying the second dose of the COVID-19 vaccine in collaboration with clinical and public health experts. While we simulate COVID-19 spread, the ideas introduced in the paper are generic and can be easily extend to other forms of agent-based simulations. Furthermore, while beyond scope of this document, DeepABM enables inverse agent-based simulations which can be used to learn physical parameters in the (micro) simulations using gradient-based optimization with large-scale real-world (macro) data. We are optimistic that the current work can have interesting implications for bringing ABM and AI communities closer.
\section{INTRODUCTION}
\label{sec:intro}

The coronavirus has had significant impact on global society and economy. Successfully navigating this phase involves answering a series of "what-if" questions that can have long lasting implications on our future. These questions may include: 'what-if' we delay the second dose of the vaccine, 'what-if' we reopen schools, 'what-if' we impose a lockdown, etc. Agent-Based Modeling has emerged as a central tool that can help policy makers ground their decisions to tackle these challenging questions. ABMs are descriptive simulation models that enable one to (i) study the actions and interactions of large heterogeneous populations, and (ii) analyze the emergent effects of behavioral and clinical interventions on various public health outcomes such as cumulative mortality, infections and hospitalizations. 

While ABMs are powerful descriptive tools, emergent behavior can be highly sensitive to the scale of the input population and calibration of the input parameters. Conventional ABM frameworks such as Mesa and NetLogo \cite{masad2015mesa,Wilensky:1999} follow an object-oriented design that is centered around agent definition and actions, which are characterized as objects. While conceptually appealing, these are often inefficient to scale to large agent populations to sufficiently represent the behavior in real-world contexts. In this work, we introduce DeepABM, a novel framework for agent-based modeling that takes a network-centric approach that revolves around the interaction networks of the agents in the simulation. DeepABM builds upon concepts of tensor-calculus and graph neural networks in deep learning to deliver scale and efficiency to these descriptive simulations. 

In DeepABM, individual agents (and their states) are modeled as tensors, while their interactions as represented as permutation-invariant message-passing operations in graph neural networks. Leveraging the advantages of and recent advances in in graph deep learning, DeepABM can seamlessly scale to large populations (of size greater than 100,000 agents) in real-time and efficiently execute on graphic processing units (GPU).
Furthermore, while not used to obtain the results discussed in this paper, DeepABM can also concurrently calibrate a large number of input parameters using supervised gradient-based optimization instead of (restrictive) randomized search methods. 


DeepABM has been adapted during the COVID-19 pandemic to efficiently simulate transmission dynamics and study the impact of various non-pharmaceutical interventions (NPIs) on various public health outcomes. We will, henceforth, refer to this model as \textit{DeepABM-COVID}. Agent distribution and their interactions in DeepABM-COVID are parameterized using real-world census data. Parameters related to disease transmission and progress are calibrated using a plethora of research studies that analyze real-life clinical data.  DeepABM-COVID models agent interaction over multiple networks (household, occupation and random) and can concurrently support several diverse interventions (quarantine, exposure notification, testing, vaccination). DeepABM-COVID can model 200 million interactions over 100,000 agents across 180 timesteps in about 90 seconds on a GPU. To contextualize the extend of this gain, we benchmarked our analysis by implementing the same simulation in Mesa, which takes around five hours to execute.


DeepABM-COVID has already been used in a research study that analyzes the impact of delaying the second dose of the two-dose COVID-19 vaccines, in favor of administering the vaccine to a broader segment of the population. This research has been recently published in British Medical Journal and has received attention from the various research communities considering effective vaccine allocation strategies. The simulation study, parameters and behavior governing vaccine efficacy, immunity against the vaccine, etc. have been designed in collaboration with leading clinical and public health experts from the Mayo Clinic and Arizona State University. Other members of the team includes computational scientists from MIT as well as biostatisticians from Mayo Clinic and Harvard University. We outline some of the relevant results from that study in Section 5 below, and refer the reader to \shortciteN{romero2021public}.

The following sections provide more detailed information about the framework and the paper is structured as: Section 2 includes preliminary for the DeepABM framework, Section 3 discussed the DeepABM-COVID framework, Section 4 highlights the various interventions that are supported. Section 5 includes a detailed case-study studying the effect of delaying second dose and conclusion is in Section 6.
\section{PRELIMINARIES}
\label{sec:deepabm}

\subsection{DeepABM}
DeepABM is a tensor calculus based approach for our simulations. In this approach, agents are not modeled as objects, but rather agent states are modeled as tensors and tensor algebra is used to model state transitions. Vectorized implementations of ABM have been explored in literature, primarily for mean-field or similar approximation based models and also for some agent based models. However, to the best of our knowledge, this is the first time a network based model that captures individual interactions, is modeled using tensor calculus. Essentially, we use the message passing abstraction of graph convolutional neural networks to model the inter-agent interaction and associated infection dynamics~\shortcite{gcn-paper}. The message passing network \cite{zhong2020hierarchical} provides infrastructure to collect messages (effects of interactions) from all neighbours of a node for all nodes in a network. This design has been central to significant progress in geometric deep learning \shortcite{gnn-egs,lecun2015deeplearning,gattnnet}. Hence, deep learning  frameworks (such as Pytorch \shortciteN{pytorch}) provide optimized differentiable implementations for this with support for GPU execution. We leverage this graph deep learning infrastructure to design interaction networks and model inter-agent interactions over them (such as disease spread) in agent-based models. The disease dynamics are modeled using standard parametrized tensor operations.

\subsection{Mathematical framework}
The underlying mathematical framework for DeepABM can be considered to be a partially observable semi-Markov game (POSMG), which is a slight modification of the definition given in \cite{littman1994markov}. Our ABM is modeled as a game as each agent receives different information (partial observations) at each step and chooses its response in a decentralized manner based on its private information and its own objective. We further consider this as a semi-Markov game as the decision epochs or state sojourn times are sampled from a distribution that depends on the current state and next state of the agent, unlike a fixed sojourn time as in Markov games. Furthermore, the action spaces of the agents are state-dependent. 
\section{COMPONENTS OF DeepABM-COVID}
\label{sec:deepabm-covid}

To design DeepABM-COVID, we adopted the basic disease infection and progression parameters used in \shortciteN{abueg2020modeling} and implemented a number of non-trivial extensions. The following sections provide information on the various components of DeepABM-COVID for the sake of completeness. Further information on choices of parameters, and other characterizations of realistic behavior can be found in \shortciteN{romero2021public}.  



\paragraph{Agent State Definition:}
At any given time step, the state of the agent is composed of static and dynamic components. The dynamic components include agent attributes that change over time when an agent interacts with others. The static components are attributes of the agent that do not change over the course of the simulation. However, these static components influence an agent's interaction with its neighborhood and the subsequent evolution of the dynamic attributes. 
This agent state is represented as a one-dimensional tensor obtained by concatenating the static and dynamic components.

For the agent representation in DeepABM-COVID, the \textit{static component} includes the following attributes: (i) age, (ii) household, (iii) occupation, (iv) (random) number of daily interactions. Each of these attributes are categorical variables and are initialised for each agent using the real-world census (for age, household, and occupation) and mobility data (to generate the number of interactions) for King's County in the State of Washington. In particular, age corresponds to an identifier from 1 to 8 (non-overlapping) age-groups and occupation corresponds to an identifier from 1 to 23 possible defined occupations. 

The \textit{dynamic component} of each agent's state includes the following attributes: (i) (current) disease stage, (ii) quarantine status, and (iii) vaccination status. At any step, the disease stage of each agent can be one of the defined eleven values: susceptible, asymptomatic, presymptomatic mild, presymptomatic severe, mild symptomatic, severe symptomatic, hospitalized, critical in ICU, recovered, vaccinated or dead. For vaccination, we assume that the available vaccines follow a two-dose regimen, and the vaccination status can take one of three distinct values: pre-vaccination (1st dose eligible), partially vaccinated (1st dose completed), and fully-vaccinated (second dose completed). The quarantine status for an agent is a binary variable: true or false. These dynamic components change over time and drive various events (such as transmission instance) in the simulations. To initialize the agent population, we use the census and mobility statistics for Kings County, WA as in \shortciteN{uwash-paper}.  

\paragraph{Interaction Networks: }
At any step of the simulation, each agent interacts with neighboring agents concurrently across three independent networks. In the DeepABM-COVID simulator, these networks are used to represent (i) \textit{household network}, which defines cross-age interactions within family members, (ii) \textit{occupation network}, which defines interactions due to the individuals' employment at their workplaces at different industries, and (iii) a \textit{random interaction network}, which defines other random interactions agents may have over the course of a day. 

There are multiple household networks in the simulation, one for every household, each of which is modeled as a fully-connected network. 
Each agent is assigned to a single household network at the initialization and this remains fixed throughout the simulation. We assume that each (non-quarantined) agent is connected to and will interact with every other agent in his household at every step.

There are 23 occupation networks (as in \shortciteN{uwash-paper}) defined in the current version of DeepABM-COVID. Every agent is assigned to a single occupation network once at initialization, and this assignment does not change throughout the simulation. Each occupation network is modeled as a small-world Watts-Strogatz network \shortcite{watts1998collective}, which is initialized once but re-parameterized at every step. What this means is that the specific agents in an occupation network (i.e., those that share the same occupation) don't change over time but each agent may interact with a different subset of agents on any given day. The re-parameterization of each occupation network is done independently in accordance with the value of the parameter that defines the mean number of daily interactions for that occupation.

There is a single global random network in the simulation. This is modeled as a small-world Watts-Strogatz network that is re-initialized at every time step. The network is used to simulate infections that result from random interactions that agents may have with unknown individuals. These interactions, for instance, may occur during visits to the grocery store, doctor's office, at the train station, etc. We observe that these random interactions play a significant role in viral spread in the simulation, in consistence with real-world estimates. At any given time step, we use each agent's (age-stratified) number of random interactions to parameterize this global random network. We define the networks using the interaction parameters given by \shortciteN{uwash-paper}. 

We next explain the transition dynamics that govern the agent state transitions over simulation steps in DeepABM-COVID. This involves studying the infection transmission that regulates per-step inter-agent interactions and the consequent inter-step per-agent disease progression. 

\paragraph{Infection Transmission: } Infection is spread through interactions between infected and susceptible individuals. The rate of transmission primarily depends upon: (i) infectiousness of the pathogen, (ii) age-dependent susceptibility of infectee to transmission, and (iii) type of interaction (i.e., in which network it occurred).The infectiousness varies over time, starting at zero when the agent is infected, peaks at an intermediate time and eventually tends to zero. Duration of infectiousness is modeled with a gamma distribution and we use the values referenced by \shortciteN{abueg2020modeling} for the mean and standard deviation as clinically studied in \shortciteN{ref-3}, \shortciteN{ref-4}, and \shortciteN{ref-5}. The infection susceptibility of an agent is modeled as age-stratified following the studies in \shortciteN{ref-6}, \shortciteN{ref-9}, and \shortciteN{ref-10}. In particular, we stratify age into nine age-groups: 0-10, 11-20, 21-30, 31-40, 41-50, 51-60, 61-70, 71-80, and 80+, and follow the susceptibility parameters as used in (\shortciteN{abueg2020modeling}). The type of interaction is defined by the specific network in which it originates. To account for the duration of interaction, which is not directly observable, we use a multiplicative scalar to amplify effect of interaction in household networks. Finally, viral transmission in each interaction is represented by the following equations:
\begin{align}
    \lambda(t, s_i, a_s, n) &= \frac{RS_{a_{s}}A_{s_{i}}B_n}{\bar{I}}\int_{t-1}^{t}f_{\Gamma}(u; \mu_i, \sigma_i^2)du,\\
    P(t, s_i, a_s, n) &= 1 - e^{-\lambda(t, s_i, a_s, n)},
\end{align}
where $t$ denotes the amount of time since infection; $s_i$ indicates the infector's symptom status (asymptomatic, mild, moderate/severe); as is the age of the susceptible; $n$ is the type of network where the interaction occurred; $I$ is the mean number of daily interactions; $f_\Gamma(u;\mu_i,\sigma_i^2)$ is the probability density function of a gamma distribution; $\mu_i$ and $\sigma_i$ are the mean and width of the infectiousness curve; $R$ scales the overall infection rate (under some simplifying assumptions it is mean number of people infected by each moderately/severely symptomatic individual); $S_{a_{s}}$ is the scale-factor for the age of the susceptible; $A_{s_{i}}$ is the scale-factor for the infector being asymptomatic; $B_{n}$ is the scale-factor for the network on which the interaction occurred.

\paragraph{Disease Progression: } When an agent is infected, it enters a hierarchy of disease progression, as summarized in \shortciteN{romero2021public}. 
This corresponds to the evolution of disease stage (which is part of the dynamic component of an agent state) of an agent during the course of the simulation. The progression time delay and probability between two disease stages primarily depends upon age of agent. 
These parameters are obtained from established clinical literature for COVID-19 disease progression. Essentially, the duration of various disease stages of an agent is randomly generated using continuous time distributions with disease stage and age-dependent parameters.

\section{MODELING OF INTERVENTIONS IN DeepABM-COVID}
The DeepABM-COVID simulator currently supports the following interventions: isolation and self-quarantine, digital exposure notification (DEN), testing, and vaccination. 

\subsection{Self-quarantine} Upon experiencing symptoms, agents undertake a diagnostic test. If the test returns positive, then the agent is self-quarantined for 14 time steps (days). There is a daily dropout probability from the quarantine to model non-compliance in the real-world. When an individual starts quarantine, the \textit{quarantine status} attribute in agent state is set to true (=1). Once an agent breaks quarantine, after 14 days or due to non-compliance, the quarantine status is reset to false. This cannot be updated until a new test is administered (following re-emergence of symptoms) which turns positive and initializes a new quarantine instance. Agent quarantine status influences the infection transmission from interactions on the given day. While agent is successfully quarantine, we scale the infectiousness of the agent to 0 when modeling interactions.

\subsection{Digital Exposure Notification (DEN)} When an agent tests positive, he starts to self-quarantine for a period of 14 days. If DEN is enabled to do so, a test notification is sent to other (non-quarantined) agents that this positive agent interacted with over the past 7 steps (days). We note that this notification is only sent to agents that also have access to the exposure notification app (kept as part of agent state variable) to mimic real-world constraints on digital contact tracing. Agents are assigned DEN app access randomly at initialization with a probability governed by an app-adoption parameter in society. The notified agents may undergo a test (based on compliance probability) on the next step of simulation and undergoes self-quarantine if the test returns positive.

\subsection{Testing} DeepABM-COVID simulations have support for various kinds of testing types: RT-PCR test (which is the most accurate type of testing available), an antigen test, and a much less accurate, rapid point-of-care test. During a specific simulation run, we assume that only one of the three tests is present in the community, although this assumption can be relaxed. Each test is parameterized by two variables: (i) specificity, and (ii) turnaround time. To model real-world constraints of sample collection, analysis and delivery, we assume the following parameters for each of the three tests: i) rapid antigen test (specificity=0.65, time=2 steps), ii) RT-PCR test (specificity=0.95, time=3 to 5 steps uniformly sampled), iii) rapid point-of-care test (specificity = 0.85, time=1 step). For all simulations, we use RT-PCR test by default, unless specified otherwise.


\subsection{Vaccination} DeepABM-COVID simulates the two-dose vaccination regimen as followed in the mRNA vaccines such as Pfizer and Moderna. In the simulation, it is possible to make different assumptions on the timing of the doses, the efficacy of each dose, and the prioritization scheme to be followed in the vaccination campaign. Further details on vaccination modeling capabilities of DeepABM-COVID can be found in Section~\ref{sec:casestudy}. 

\subsection{Sample Simulation Results with Interventions}
\begin{figure*}
    \centering
    \includegraphics[scale=0.2]{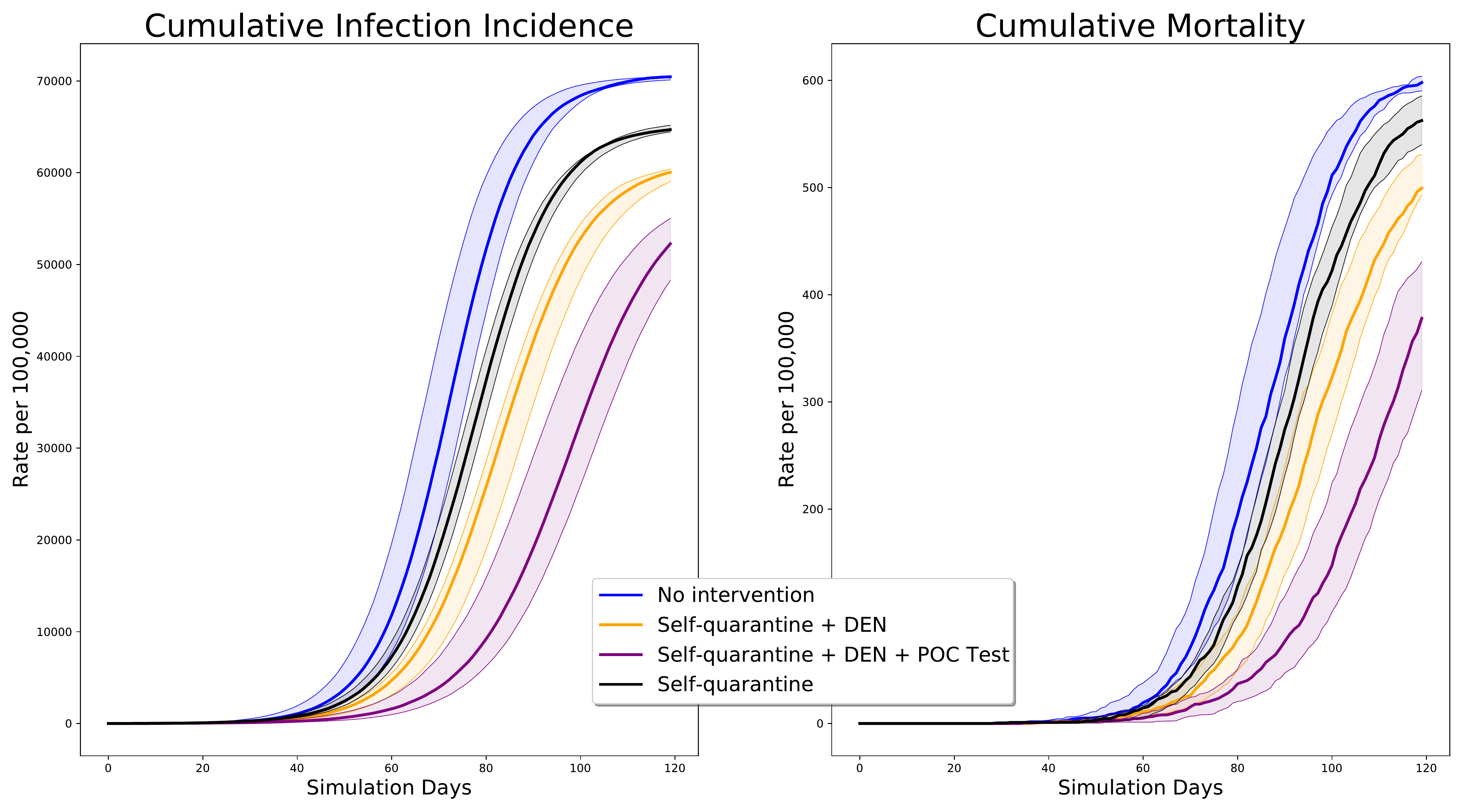}
    \caption{Mean and standard deviation for cumulative incidence of mortality and infections when simulation with different combinations of interventions. POC Test refers to the Rapid Point-of-care Testing intervention.}
    \label{fig:sample-result}
\end{figure*}

We study the effect of simulating with the different interventions presented above and present few sample results in Figure~\ref{fig:sample-result}. The results correspond to mean and standard deviation of 10 independent runs over 120 steps. We compare four configurations: a) no-intervention b) self-quarantine c) self-quarantine + DEN d) self-quarantine + DEN + Rapid Point-of-Care Testing. To ensure consistence with real-world scenarios, we assume DEN app-adoption of 0.3 and an app compliance probability of 0.8. The point of care test has turnaround time of 1 step (i.e. same day) and we assume specificity of 0.85 to account for real-world variability in sample collection and analysis. The results show that all interventions help reduce cumulative incidence of deaths and infections and incremental addition of new interventions further helps in controlling spread. We study the effect of the vaccination intervention with detailed sensitivity analysis in Section 5.

\section{CASE STUDY: IMPACT OF DELAYING SECOND DOSE OF COVID-19 VACCINES}
\label{sec:casestudy}

One of the most significant accomplishments of the COVID-19 pandemic response has been the development, manufacture and deployment of vaccines against the disease. Massive investment and innovative science have yielded multiple effective SARS-CoV-2 vaccines in record time. Two of the most effective vaccines, namely those produced by Pfizer and Moderna, consist of two-dose regimens with the second dose administered 21 or 28 days following the initial dose, while the approved viral vector vaccine received approval with a second dose 4-12 weeks after the first. As with most clinical trials of new therapeutics, dosing regimens were decided a priori, based on a combination of preliminary data and intuition. Emerging data, however, have suggested that most of the protective benefit conferred by SARS-CoV-2 vaccines may result from a single dose alone. If true, this may have significant implications regarding the optimal allocation of limited vaccines. In particular, multiple public health authorities have speculated that deviating from the standard two-dose regimen in favor of a broader one-dose regimen may save lives and reduce infectious spread. 

Despite the increasing pace of vaccinations in the US, the vaccine supply, relative to the worldwide populations in need remains constrained. We therefore sought to understand the population health implications of delaying the administration of second dose vaccines in favor of vaccinating a broader number of people with a first dose. In particular, our goal in this study was to compare the effectiveness of the \textit{standard dosing} strategy to that of a \textit{delayed second dose} strategy, which prioritizes the vaccination of a first-dose eligible individual using an agent-based simulation model. In alignment with the age-based prioritization, both strategies we considered observe prioritization by age. Despite the fact that our agent based simulation model involves representations of social and employment networks, we did not consider any prioritization with respect to employment sector; i.e., healthcare workers, for example, were not prioritized over other employment networks. In addition to the standard dosing and delayed second dose strategies, we consider a third alternative strategy, which we refer to as \textit{delayed second dose except for 65+} to test if the increased mortality risk among the elderly implies a differential vaccine dosing regimen for this segment of the population. 

It is useful to provide the following example to demonstrate the three vaccination strategies clearly, we offer the following example of six hypothetical individuals: Adam - first dose eligible 78 yr old; Betty - second dose eligible 78 yr old; Charlie - first dose eligible 68 year old; David - second dose eligible 68 year old; Eleanor - first dose eligible 40 year old; Frank - second dose eligible 40 year old. Table~\ref{tab:strategies} below shows the order in which these individuals are prioritized for vaccine administration under each strategy.

\begin{table}[h]
\caption{Description of the three strategies to be compared}
\label{tab:strategies}
\small \centering
\begin{tabular}{|l|l|l|}
\hline
Name of Strategy                                                                   & Description                                                                                                                                                                         & \begin{tabular}[c]{@{}l@{}}Priority list under \\ the strategy will be\end{tabular}    \\ \hline \hline
\begin{tabular}[c]{@{}l@{}}Standard dosing \\ (two doses on schedule)\end{tabular} & \begin{tabular}[c]{@{}l@{}}Age prioritized vaccination with second dose \\ at 21 days\end{tabular}                                                                                  & \begin{tabular}[c]{@{}l@{}}Betty, David, Frank, \\ Adam, Charlie, Eleanor\end{tabular} \\ \hline
Delayed second dose                                                                & \begin{tabular}[c]{@{}l@{}}Age prioritized vaccination prioritizing \\ vaccination of first-dose eligible individuals\end{tabular}                                                  & \begin{tabular}[c]{@{}l@{}}Adam, Charlie, Eleanor, \\ Betty, David, Frank\end{tabular} \\ \hline
\begin{tabular}[c]{@{}l@{}}Delayed second dose \\ except for 65+\end{tabular}      & \begin{tabular}[c]{@{}l@{}}Age prioritized vaccination with first dose \\ prioritization, except for the elderly (above 65), \\ who received a second dose on schedule\end{tabular} & \begin{tabular}[c]{@{}l@{}}Adam, Betty, Charlie, \\ David, Eleanor, Frank\end{tabular} \\ \hline
\end{tabular}

\end{table}

To analyze the effectiveness of these three strategies, we use the DeepABM-COVID simulator explained above. The modular design of the DeepABM toolkit enabled seamless implementation of various prioritization logic with minimal edits. Further, we also augment several extensions to accurately reflect the immunity that first dose and second dose vaccines confer on individuals, as we explain below. Various further details on the statistical analysis of clinical data to justify our choices of these parameters are provided in \shortciteN{romero2021public}.



We assume that a dose of the vaccine provides a certain probability of becoming immune to infections. This probability is dependent on whether the vaccine is administered as a first dose or a second dose. In accordance with trial data, we assume that twelve days after receiving the first dose of the vaccine, agents have a 60\%, 70\%, 80\% or 90\% probability of becoming immune (depending on the scenario). After receiving the second dose, agents reach a 95\% probability of becoming immune. 

We consider two versions of the immunity offered by the vaccine. In the first version, we assume that the vaccine provides sterilizing immunity, meaning that agents, even when they are exposed to the virus, do not get infected. Under the other version of the immunity effect we tested, agents still get infected at the same rate as non-vaccinated agents, but they experience asymptomatic infections with the associated probabilities. In this case, they can still transmit the disease with the same probability as a non-vaccinated asymptomatic patient.

Under each vaccination strategy, we run 15 replications of the agent based simulation for 180 days, and observe the number of infections, hospitalizations and deaths over time for each simulation replication. We initialize the simulations with 10 infected agents and start vaccinations when the number of infected agents reach 1\% of the population, which happens around simulated day 20. We plot the median as well as 25th and 75th percentiles of those 15 runs for each outcome of interest under the vaccination strategies to compare and contrast them with respect to their effectiveness in reducing the number of cumulative deaths as well as cumulative infections and the number of hospitalizations over time. We evaluate the performance of the three strategies under a number of different carefully constructed cases to support decisions by public health officials tasked to design and deploy vaccination campaigns. More details can be found in \shortciteN{romero2021public} but we include below two important observations on the impact of (i) efficacy of the first dose of vaccine and (ii) daily vaccine administration rate.

\noindent \paragraph{Impact of Vaccine Efficacy on the Effectiveness of Delayed Second Dose Strategy:} To evaluate the effectiveness of delaying the second dose, we first set the daily vaccination rate to 0.3\% of the population size (100K in our simulation study) and observe the differences in the cumulative deaths under each vaccination strategy. The vaccine administration rate of 0.3\% per day is obtained from the reported vaccination numbers in the U.S. (early in the vaccination campaign) and other countries around the world. At the time of writing of this article, the US vaccination rates have increased dramatically, and consequently, we analyzed the effectiveness of these strategies under higher daily vaccination rate capability, as we describe below. 

\begin{figure}[htb]
{\centering
\includegraphics[width=0.95\textwidth]{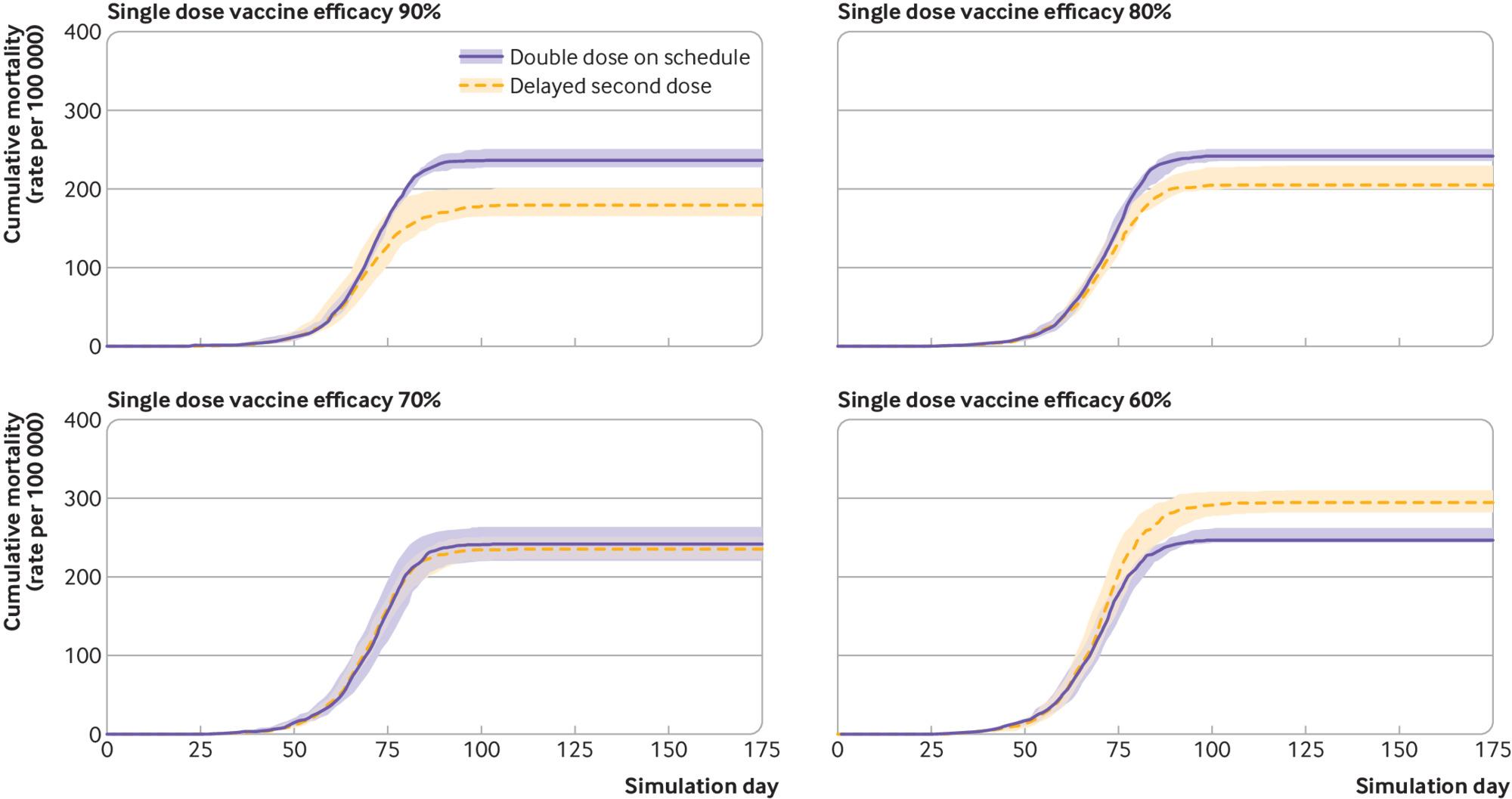}
\caption{Comparison of cumulative mortality for delayed second dose versus standard dosing under four different first dose vaccine efficacy assumptions. 
Adopted from Santiago Romero-Brufau et al. BMJ 2021;373:bmj.n1087, copyright 2021 British Medical Journal Publishing Group. \label{fig:csfig1}}
}
\end{figure}

Figure~\ref{fig:csfig1} is adopted from \shortciteN{romero2021public}, and provides the quartiles of observed cumulative deaths in the replications under the standard dosing and delayed second dose strategies for a daily vaccination rate of 0.3\%. The results demonstrate the important finding that the comparative effectiveness of the delayed second dose strategy depends strongly on the efficacy of the first dose vaccine. In particular,the total cumulative mortality on day 180 is lower for the delayed second dose scenario under the assumption that the first dose effectiveness is higher than 80\%, which is typically justified by the data from the clinical trials as well as vaccine deployment programs all over the world for the two-dose Pfizer and Moderna vaccines. 

\noindent \paragraph{Impact of Vaccine Administration Rate on Effectiveness of Delayed Second Dose Strategy:} The above presented results are also dependent on the speed at which the vaccination campaign can be run. To test this, we adopt a first dose vaccine efficacy of 80\% (which is well supported by the studies on clinical and trial data) and evaluate the effectiveness of the delayed second dose strategy under daily administration rates of 0.1\% (very slow rate), 0.3\% (nominal rate) and 1\% (relatively fast rate reflective of more recent US vaccination rates). 

\begin{figure}[htb]
{\centering
\includegraphics[width=0.5\textwidth]{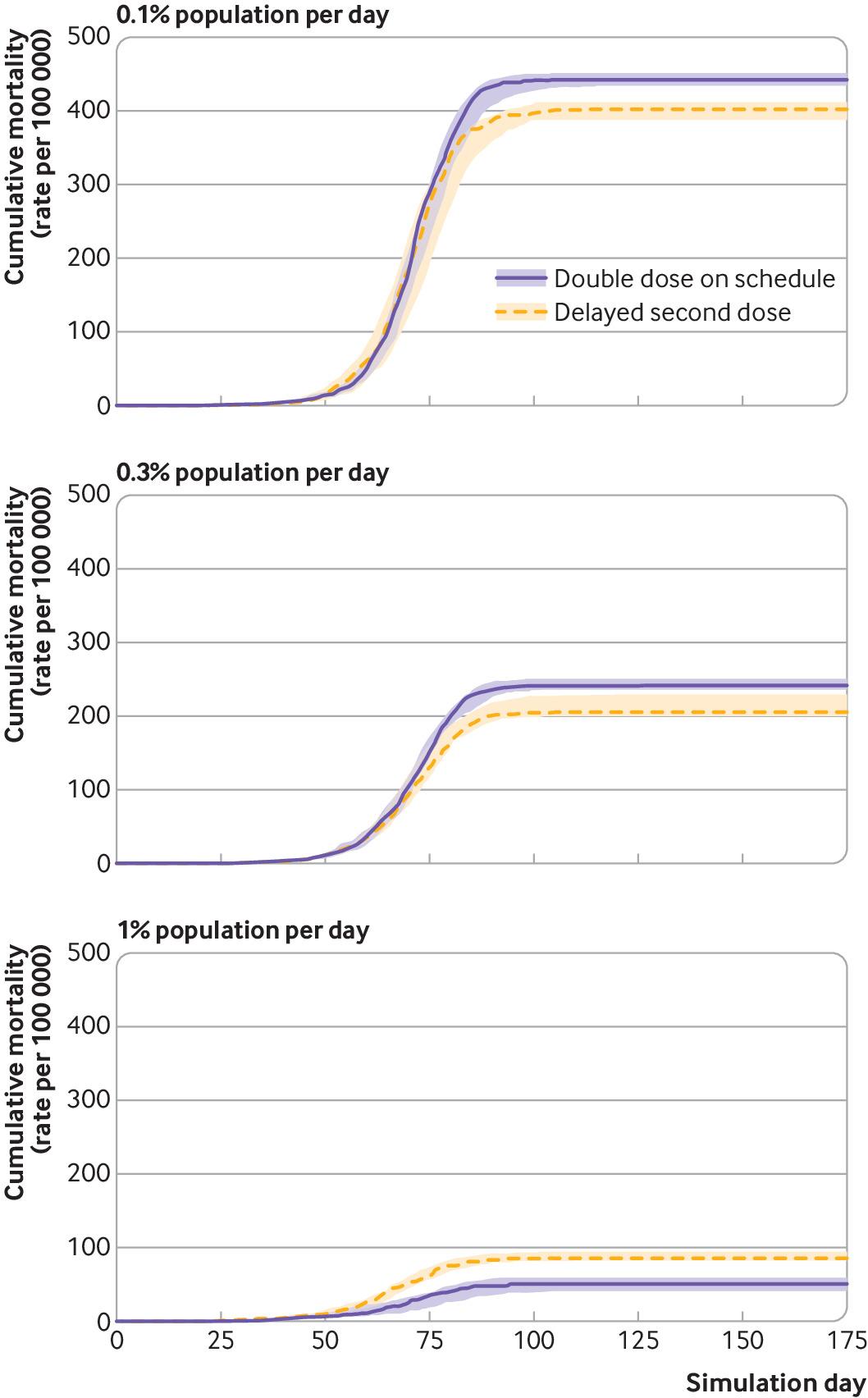}
\caption{Comparison of cumulative mortality for delayed second dose versus standard vaccination strategy under three different vaccination rate assumptions. 
Adopted from Santiago Romero-Brufau et al. BMJ 2021;373:bmj.n1087, copyright 2021 British Medical Journal Publishing Group. \label{fig:csfig2}}
}
\end{figure}

Figure~\ref{fig:csfig2} is again adopted from \shortciteN{romero2021public}, and demonstrates the above cited effect clearly.  Essentially, the standard dosing becomes the preferable strategy as the daily administration rates improve. However, given that most countries are lagging behind in vaccination efforts, this study points to the important advantage that delaying second doses offers by providing a broader, albeit less protective, first-dose administration across the population. 

The case study clearly demonstrates the impact of a computationally efficient simulation framework like DeepABM-COVID. At each stage of the pandemic, there are interesting public health policy questions that our team of experts plan to explore using the modeling and simulation capabilities offered by DeepABM-COVID. The DeepABM-COVID simulator is also made available online, to help researchers with modeling various interventions for COVID as well as adapting the same for studying other interesting emergent phenomenon in public health and beyond.
\section{CONCLUSIONS}
\label{sec:conclusion}

In this paper, we introduce DeepABM, a toolkit for agent-based modeling that leverages graph neural network frameworks from deep learning to bring scale and efficiency to agent-based simulations. DeepABM can seamlessly scale to large populations (with more than 100,00 agents) in real-time and also execute efficiently on a GPU. We extend the toolkit to introduce \textit{DeepABM-COVID} for simulating spread of COVID-19 with (concurrent) support for several interventions. We use DeepABM-COVID to specifically study the public health impact In particular, we present a sample of our results on delaying second dose of the mRNA vaccine and present recommendations on when this strategy could be usefully adopted. 

An interesting direction of future research is to leverage DeepABM to introduce \textit{inverse agent-based simulations} which can be used to learn physical parameters in the (micro) simulations using gradient-based optimization with large-scale real-world (macro) data. This can be used to calibrate agent-based simulations, as a significant shift from current grid search techniques, as well as enable them for real-world predictive modeling. Furthermore, DeepABM separates modeling of agent transition and agent behavior, enabling learning of adaptive behavior of agents by searching over a space of rules instead of using fixed rule-based behavior. We are optimistic that the current work can have interesting implications for bringing ABM and AI communities closer.

\footnotesize

\bibliographystyle{wsc}

\bibliography{demobib}

\section*{AUTHOR BIOGRAPHIES}

\noindent {\bf AYUSH CHOPRA} is a computer scientist and researcher at the Massachusetts Institute of Technology (MIT) and a graduate student at the MIT Media Laboratory. 
His research focuses on machine learning in decentralized (single and multi-agent) systems with applications in public health and economics. He previously worked as a researcher for the Media and Data Science Research Lab at Adobe. His work has been published at several top-tier ML conferences and interdisciplinary journals, and resulted in 23 patents. 
Ayush received his B.Tech in Computer Science from Delhi Technological University in India. His email address is \email{ayushc@mit.edu}. \\

\noindent {\bf ESMA S. GEL, PhD} is an Associate Professor of Industrial Engineering at the School of Computing, Informatics and Decision Systems Engineering of Arizona State University. Gel's research focuses on the use of stochastic modeling and control techniques for the design, control and management of operations. 
Gel holds 
M.S. and Ph.D. degrees in Industrial Engineering from Northwestern University, obtained in 1995 and 1999, respectively.  Her email address is \email{esma.gel@asu.edu}. \\

\noindent {\bf JAYAKUMAR SUBRAMANIAN} is a Senior Research Scientist at the Media and Data Science Research Lab at Adobe, India. His research interests include reinforcement learning in single and multi-agent systems. He has a Ph.D. in reinforcement learning in partially observed and multi-agent systems from McGill University and dual degrees (Bachelor + Master) in Aerospace Engineering from the Indian Institute of Technology, Bombay. His email address is \email{jasubram@adobe.com}.\\

\noindent {\bf BALAJI KRISHNAMURTHY} is a Principal Scientist and Director of Media and Data Sciences Research Lab at Adobe, India. His research interests span Computer Vision and Machine Learning applied to large scale systems for marketing, personalization and decision support His email address is \email{kbalaji@adobe.com}.\\

\noindent {\bf SANTIAGO ROMERO-BRUFAU, MD, PhD} is Assistant Professor of Medicine and Healthcare Systems Engineering at Mayo Clinic, where he also serves as Principal Data Scientist for the Department of Medicine. His research interest focuses on the development of clinical decision support tools using machine learning and data science, and in the implementation of these tools into clinical workflows to improve the quality and efficiency of care. His email address is \email{santiagoromerobrufau@hsph.harvard.edu}.\\

\noindent {\bf KALYAN S. PASUPATHY, PhD} is an expert in systems science and health informatics and is focused on both advancing the science and translating knowledge to improve care delivery, demonstrated through his academic and practice leadership roles. Dr. Pasupathy is the founding scientific director of the Mayo Clinic Clinical Engineering Learning Laboratories. 
His email address is \email{Pasupathy.Kalyan@mayo.edu}.\\

\noindent {\bf THOMAS C. KINGSLEY, MD} is a physician and Assistant Professor of Medicine at Mayo Clinic and is the Chief Epidemiologist and Health Officer for Path Check Foundation. 
He is an applied informatics and AI researcher with the goal of translating data to model to innovation or improved decision making and policy. He has won multiple awards including best AI project of 2020 at Mayo Clinic. 
His email address is \email{Kingsley.Thomas@mayo.edu}. \\

\noindent {\bf RAMESH RASKAR, Ph.D.} is an Associate Professor at MIT Media Lab and directs the Camera Culture research group. His focus is on AI and Imaging for health and sustainability. He is also founder and chairman at PathCheck foundation, a non-profit for COVID-19 response and has deployed digital contact tracing solutions in several states. He received the Lemelson Award (2016), ACM SIGGRAPH Achievement Award (2017), DARPA Young Faculty Award (2009), Alfred P. Sloan Research Fellowship (2009), TR100 Award from MIT Technology Review (2004) and Global Indus Technovator Award (2003). His email address is \email{raskar@mit.edu}.

\end{document}